\documentclass{PoS}

\usepackage{epstopdf}
\DeclareGraphicsExtensions{.png, .jpg, .pdf, .eps}
\usepackage{aas_macros}

\def\msun{$M_{\odot}$}

\def\ergsec{\hbox{erg s$^{-1}$}}

\def\Herschel{\emph{Herschel}}

\title{\Herschel\ views on ultra-luminous X-ray sources}

\ShortTitle{\Herschel\ views on ultra-luminous X-ray sources}

\author{\speaker{Mathieu Servillat}, Alexis Coleiro\\
        Laboratoire AIM (CEA/DSM/IRFU/SAp, CNRS, Universit\'e Paris Diderot), CEA Saclay, Bat.~709, 91191 Gif-sur-Yvette, France\\
        E-mail: \email{mathieu.servillat@cea.fr}}

\author{Sylvain Chaty\\
        Laboratoire AIM (CEA/DSM/IRFU/SAp, CNRS, Universit\'e Paris Diderot), CEA Saclay, Bat.~709, 91191 Gif-sur-Yvette, France\\
        Institut Universitaire de France, 103 bd Saint-Michel, 75005 Paris, France}

\abstract{
The nature of ultra-luminous X-ray sources (ULXs), which are off-nuclear extragalactic X-ray sources that exceed the Eddington luminosity for a stellar-mass black hole, is still largely unknown. They might be black hole X-ray binaries in a super-Eddington accretion state, possibly with significant beaming of their emission, or they might harbor a black hole of intermediate mass ($10^{2}$ to $10^5$ solar masses). Due to the enormous amount of energy radiated, ULXs can have strong interactions with their environment, particularly if the emission is not beamed and if they host a massive black hole.

We present early results of a project that uses archival \Herschel\ infrared observations of galaxies hosting bright ULXs in order to constrain the nature of the environment surrounding the ULXs and possible interactions. We already observe a spatial correlation between ULXs and dense clouds of cold material, that will be quantified in subsequent work.
Those observations will allow us to test the similarities with the environment of Galactic high mass X-ray binaries. This project will also shed light on the nature of the host galaxies, and the possible factors that could favor the presence of a ULX in a galaxy.
}

\FullConference{ An INTEGRAL view of the high-energy sky (the first 10 years) - 9th INTEGRAL Workshop and celebration of the 10th anniversary of the launch\\
                 15-19 October 2012\\
                 Bibliotheque Nationale de France, Paris, France}

\begin{document}

\section{Introduction}

Ultra-luminous X-ray sources (ULXs) are non-nuclear extragalactic objects with bolometric luminosities $>10^{39}$~\ergsec\ (e.g. \cite{Roberts:2007p3209}). Luminosities up to $10^{41}$~\ergsec\ can be plausibly explained through beaming effects \cite{King:2008p10770,Freeland:2006p4898} and/or hyper-accretion onto stellar mass black holes (BHs) \cite{King:2008p10770,Begelman:2002p4262}.
A rare class of ULXs, the hyper-luminous X-ray sources, have X-ray luminosities $>10^{41}$~\ergsec\ and require increasingly complicated and unlikely scenarios to explain them without invoking the presence of a more massive BH.

Two varieties of BHs have strong observational evidence to date: stellar mass (3--20~\msun) BHs and supermassive BHs ($10^{6-9}$~\msun) present in the cores of most galaxies. It is believed that stellar mass BHs are formed from the collapse of massive stars (e.g. \cite{Fryer:2003p10778}), but it is not yet clear how supermassive ones are formed. One model proposes that they are formed from the mergers of smaller mass BHs ($10^2$ to $10^5$~\msun), the so-called intermediate mass BHs (IMBHs, e.g. \cite{Madau:2001p10784}).
To date, there are very few observational evidence for IMBHs. One of the best candidate is the most luminous ULX discovered in the galaxy ESO 243-49 dubbed HLX-1 \cite{Farrell:2009p1164,Servillat:2011p6563,Farrell:2012p7818}.
A cutoff in the luminosity function of X-ray sources is found around $2 \times 10^{40}$~\ergsec\ (e.g. \cite{Swartz:2011p8275}), suggesting that low luminosity ULXs are connected to high mass X-ray binaries, while the most luminous could represent another kind of population with more massive BHs.

\section{Motivations and objectives}

Here, we propose to test the connection between ULXs and gas/dust clouds in their host galaxy, that are related to star forming regions. The \Herschel\ observatory can be used to map the distribution of cold gas and dust in galaxies in an unprecedented way, with diffraction limited resolution and high sensitivity in the mid-far infrared (70 to 670 $\mu$m).
A large number of galaxies hosting ULXs has now been imaged with \Herschel.

The main objective is to test the spatial correlation between ULXs and clouds of cold material in the galaxies. Using \Herschel\ measurements we can access the characteristics of the cold gas and molecular clouds such as temperature and density. The gas/dust distribution can then be compared to the young stellar population distribution revealed e.g. in ultraviolet maps to check the relation between ULXs and star forming regions. This will bring further clues on the origin of ULXs.

Moreover, ULXs radiate an enormous amount of energy and are expected to have strong interactions with their environment. The soft X-rays could be absorbed by surrounding gas/dust and re-emitted isotropically in the mid-far infrared. A point like source connected to a ULX would provide a unique way to test the amount of energy radiated by the ULX over time, and to test the isotropy of its X-ray emission. A bright ULX emitting isotropically is likely to host an IMBH.

\section{Selection of ULXs}

\begin{figure}
\centering
\includegraphics[width=.9\columnwidth]{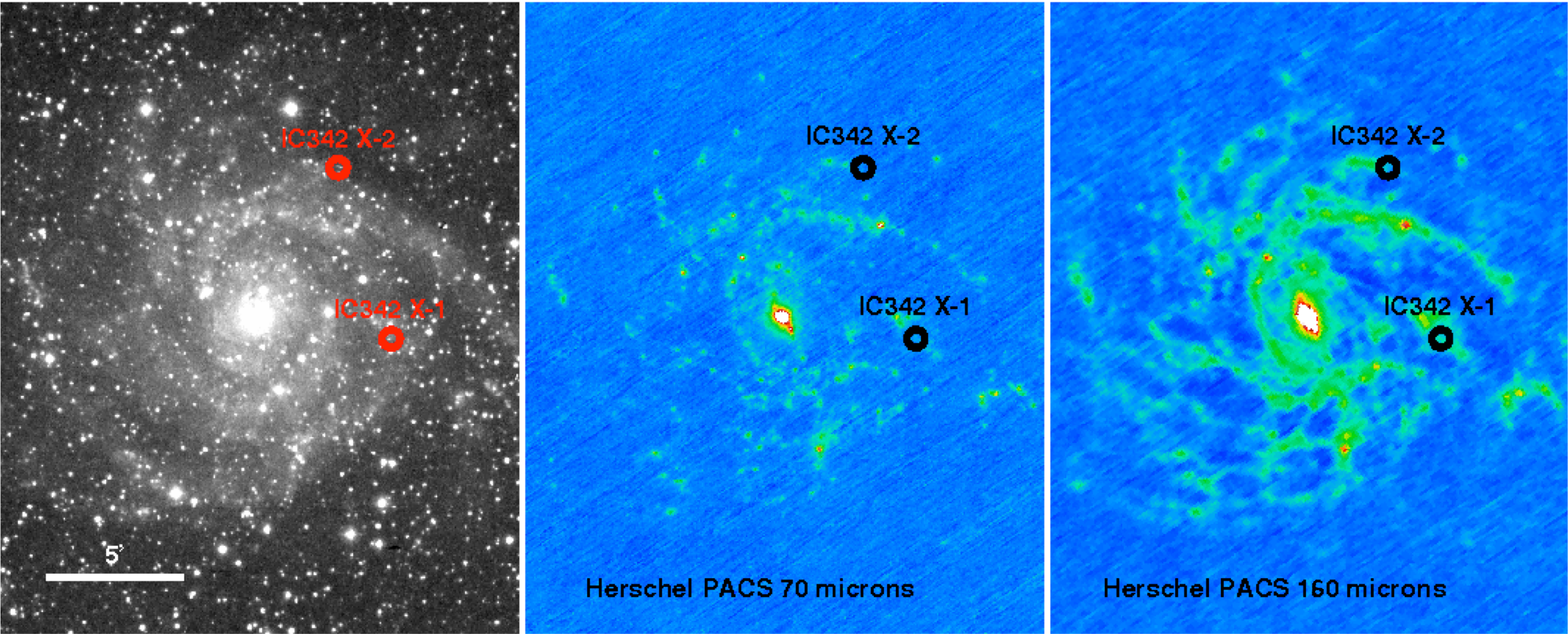}
\caption{IC 342, spiral galaxy at 3.3$\pm$0.3~Mpc, with 2 ULXs, X-1: $L_X = 3.6 \times 10^{40}$~\ergsec\ and X-2: $L_X = 1.1 \times 10^{40}$~\ergsec\ as seen with SDSS and \Herschel\ (70 and 160 $\mu$m), from left to right.}
\label{fig:ic342}
\end{figure}

\begin{figure}
\centering
\includegraphics[width=.9\columnwidth]{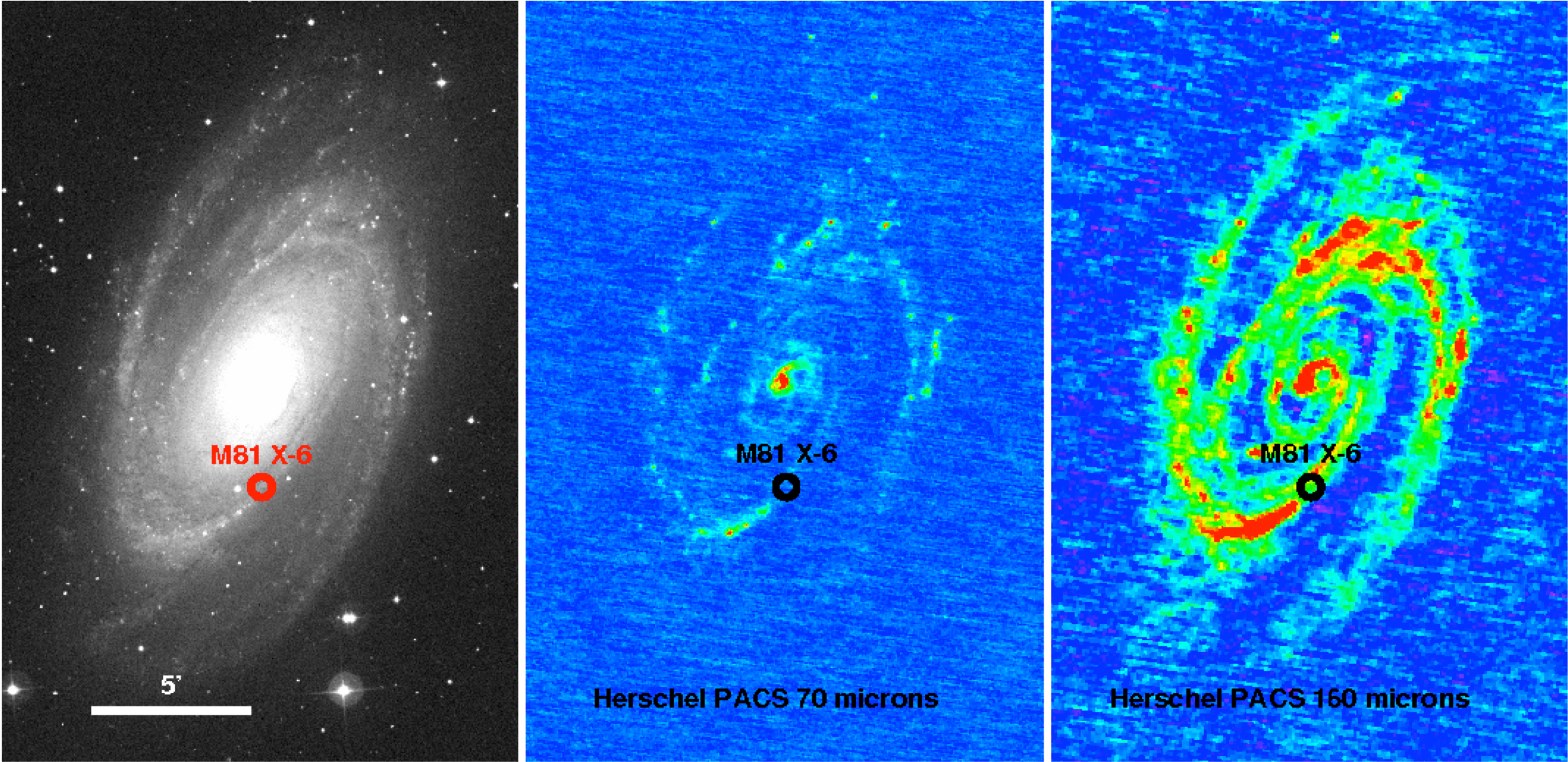}
\caption{M81, spiral galaxy at 3.6$\pm$0.12~Mpc, with the ULX X-6 ($L_X = 5.4 \times 10^{39}$~\ergsec) as seen with SDSS and \Herschel\ (70 and 160 $\mu$m), from left to right.}
\label{fig:m81}
\end{figure}

\begin{figure}
\centering
\includegraphics[width=.9\columnwidth]{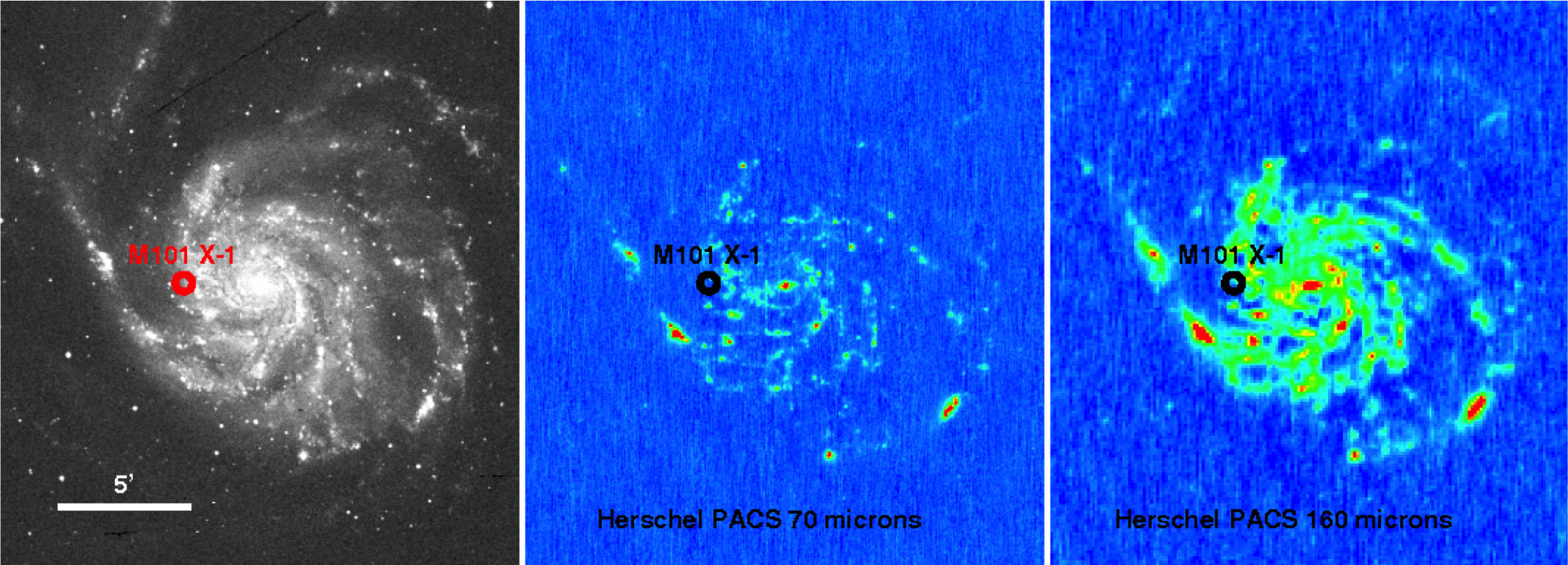}
\caption{M101, spiral galaxy at $\sim$7~Mpc, with the ULX X-1 ($L_X = 3.6 \times 10^{40}$~\ergsec) as seen with SDSS and \Herschel\ (70 and 160 $\mu$m), from left to right.}
\label{fig:m101}
\end{figure}

\begin{figure}
\centering
\includegraphics[width=.9\columnwidth]{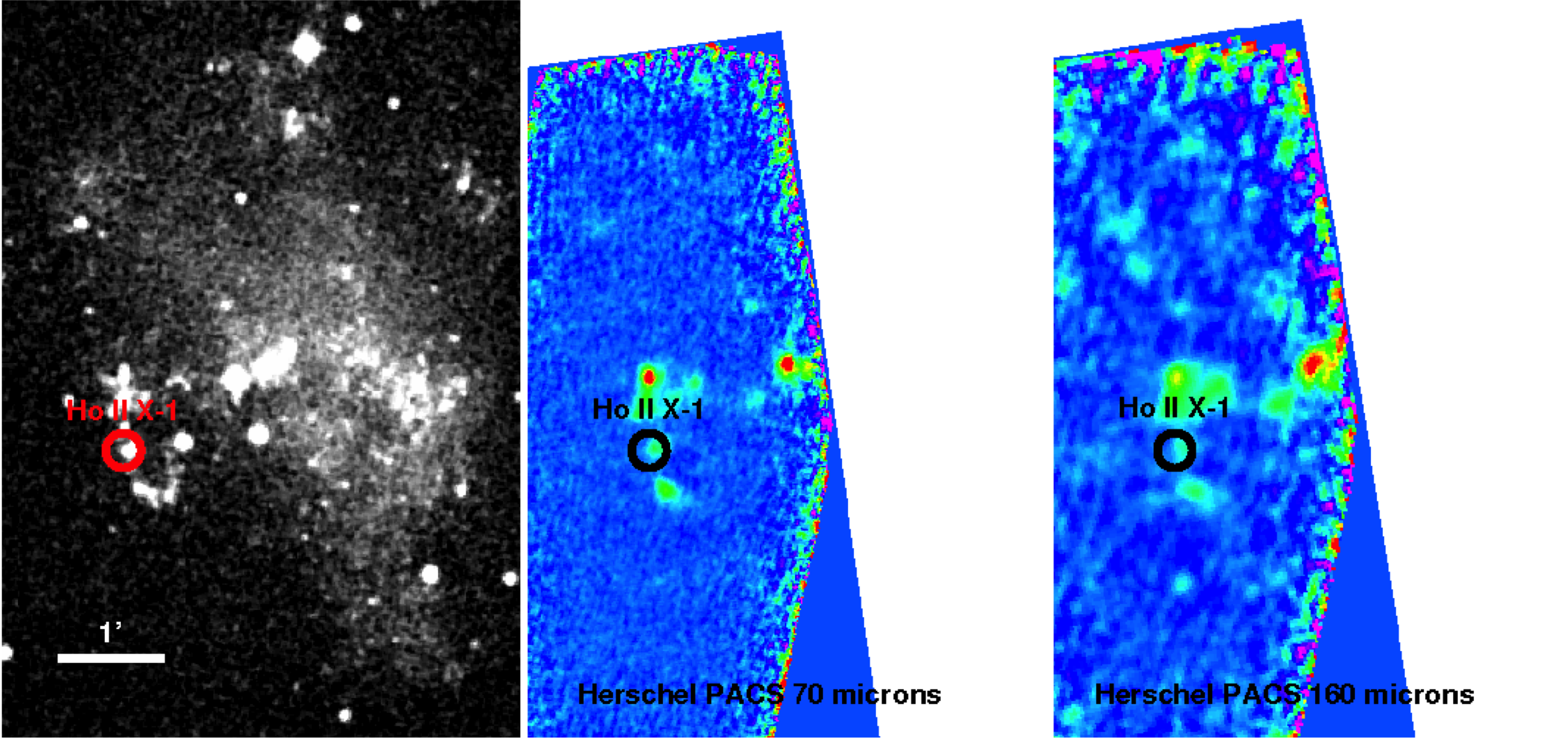}
\caption{Holmberg II, dwarf irregular galaxy near M81 at 3.6$\pm$0.12~Mpc, with the ULX X-1 ($L_X = 1.4 \times 10^{40}$~\ergsec) as seen with SDSS and \Herschel\ (70 and 160 $\mu$m), from left to right.}
\label{fig:HoII}
\end{figure}

We used the most complete and recent catalogs of ULXs detected with Chandra \cite{Swartz:2011p8275} and XMM-Newton \cite{Walton:2011p8388,Sutton:2012p9945} and focused primarily on the brightest ($>10^{40}$~\ergsec) and closest ($<40$~Mpc) ULXs. We then checked the \Herschel\ Science Archive for images and found data for a sample of 20 host galaxies containing ULXs (examples of images are shown Figures~\ref{fig:ic342}, \ref{fig:m81}, \ref{fig:m101}, and~\ref{fig:HoII}).

We are using HIPE \cite{Ott:2010p10804} to reprocess the data with custom scripts for PACS (M. Sauvage, private communication) and SPIA for SPIRE data \cite{Schulz:2011p10849}. The software getsources \cite{Menshchikov:2012p9675} can then be used to detect the sources and filaments, and to extract fluxes with a multi-wavelength method.

\section{Discussion and preliminary conclusions}

In the selection of images presented here, we can already see a correlation between ULXs and molecular clouds, e.g. in the arms of spiral galaxies, that we will quantify in subsequent work. This would suggest a strong link with star forming regions.
However, an interesting new result is that ULXs seem to be slightly offset from the brightest features. We plan to estimate the typical offset, which can then be used to derived an approximate age for the ULXs, as done in \cite{Coleiro:2013p10853}.

For ULXs possibly associated with a far-infrared source, such as HoII X-1 in Figure~\ref{fig:HoII}, the soft X-rays may be absorbed re-emitted isotropically in the infrared, so it would be possible to determine whether the X-ray emission is isotropic or beamed. Hence we may distinguish ULXs that are sub- or super-Eddington emitters, and get constraints on the mass of the BH they harbor.

We will also compare those observations to dedicated observations of Galactic high mass X-ray binaries with \Herschel, that confirmed the presence of dust in some of those objects, and enabled the study of absorbing material, either enshrouding the whole binary systems, or surrounding the sources within their close environment (see Chaty et al. in those proceedings).

Finally, we will be able to isolate some characteristics that could favor the presence of a ULX in a galaxy, such as disturbance in the gas/dust distribution, mass and temperature of the cold material, or the type of galaxies.


\acknowledgments

MS acknowledges founding from the Centre National d'Etudes Spatiales (CNES), and we thank the CNES for supporting MINE (the Multi-wavelength INTEGRAL NEtwork).


\bibliographystyle{JHEP} 
\bibliography{../../../../Work/papers/ref.bib}

\end{document}